\documentclass{aa}  

\usepackage{graphicx}
\usepackage[dvipsnames]{color}
\usepackage{ulem}
\usepackage{txfonts}
\newcommand{\LCDM}{$\Lambda$CDM}
\newcommand{\kms}{$\mathrm{km} \, \mathrm{s}^{-1}$}

\newcommand{\VhaloDM}{$V_\mathrm{max}^\mathrm{DM}$}
\newcommand{\VhaloHydro}{$V_\mathrm{max}^\mathrm{hydro}$}

\newcommand{\Vrot}{$V_\mathrm{rot,HI}$}

\newcommand{\Rout}{$R_\mathrm{out,HI}$}
\newcommand{\rout}{$R_\mathrm{out}$}
\newcommand{\Vout}{$V_\mathrm{out,HI}$}

\begin{document}

\title{Testing core creation in hydrodynamical simulations using the HI kinematics of field dwarfs}

\author{
E. Papastergis\inst{\ref{kapteyn}}\fnmsep\thanks{\textit{NOVA} postdoctoral fellow}, 
\and A. A. Ponomareva\inst{\ref{anu},\ref{kapteyn}}
}

\institute{
Kapteyn Astronomical Institute, University of Groningen, Landleven 12, Groningen NL-9747AD, The Netherlands \\ \email{papastergis@astro.rug.nl,ponomareva@astro.rug.nl}\label{kapteyn}
\and
Research School of Astronomy \& Astrophysics, Australian National University, Canberra, ACT 2611, Australia \label{anu}
}

\titlerunning{Core creation vs. HI kinematics}


\abstract{
The majority of recent hydrodynamical simulations indicate the creation of central cores 
in the mass profiles of low-mass halos, a process that is attributed to star formation-related baryonic feedback. Core creation is regarded as one of the most promising solutions to potential issues faced by lambda cold dark matter (\LCDM) 
cosmology on small scales. For example, the reduced dynamical mass enclosed by cores can explain the low rotational velocities measured for nearby dwarf galaxies, thus possibly lifting the seeming contradiction with the \LCDM \ expectations (the so-called ``too big to fail'' problem). Here we test core creation as a solution of cosmological issues by using a sample of dwarfs with measurements of their atomic hydrogen (HI) kinematics extending to large radii. Using the NIHAO hydrodynamical simulation as an example, we show that core creation can successfully reproduce the kinematics of dwarfs with small kinematic radii, $R \lesssim 1.5$ kpc. However, the agreement with observations becomes poor once galaxies with kinematic measurements extending beyond the core region, $R \approx 1.5 - 4$ kpc, are considered. This result illustrates the importance of testing the predictions of hydrodynamical simulations that are relevant for cosmology against a broad range of observational samples. We would like to stress that our result is valid only under the following set of assumptions: i) that our sample of dwarfs with HI kinematics is representative of the overall population of field dwarfs, ii) that there are no severe measurement biases in the observational parameters of our HI dwarfs (e.g., related to inclination estimates), and iii) that the HI velocity fields of dwarfs are regular enough to allow the recovery of the true enclosed dynamical mass.
}


   \maketitle
%

\section{Introduction}
\label{sec:intro}

Dissipationless simulations in the lambda cold dark matter (\LCDM) \ cosmology have consistently produced halos with a steep (or ``cuspy'') inner density profile \citep[e.g.,][]{NFW1997}. On the other hand, the majority of recent hydrodynamical simulations --which are able to incorporate the effects of baryonic processes-- predict the creation of low-density cores in the inner regions of small halos (\citealp[to name a few]{Mashchenko2008,Governato2010,Zolotov2012,Madau2014,Onorbe2015,Tollet2016}; but see also \citealp{Sawala2016}). Core creation is theoretically attributed to repeated episodes of gas blow-out from the halo center, following bursts of star formation activity \citep{PontzenGovernato2012}.

Core creation has been identified as one of the most promising solutions to long-standing issues of \LCDM \ cosmology on the scale of dwarf galaxies. In addition to its obvious potential to resolve the ``cusp vs. core'' controversy \citep[e.g.,][]{Oh2011b,Read2016,Katz2016}, core creation has also been proposed as a solution to the related too big to fail (TBTF)  cosmological problem \citep{Boylan2011}. In essence, the TBTF problem refers to the fact that it is very difficult to reproduce both the number density of dwarfs and their internal kinematics within the \LCDM \ context. More specifically, the paucity of low-mass dwarfs observed in galaxy surveys \citep{Papastergis2011,Kelvin2014,Klypin2015} implies that their host halos are relatively massive, as higher mass halos are much less abundant than lower mass halos in \LCDM \ \citep[e.g.,][]{BolshoiP}. However, the internal rotational velocities measured in dwarfs \citep[e.g.,][]{Kirby2014} are too low to be compatible with such a host assignment. Core creation can potentially resolve the issue as the lower-than-expected rotational velocities of dwarfs can be explained by the diminished enclosed dynamical mass of a cored density profile \citep{BrooksZolotov2014,Madau2014,BrookdiCintio2015a,Chan2015}. This effect has been shown most explicitly by \citet{Dutton2016}, who have used the NIHAO hydrodynamical simulation of galaxy formation \citep{Wang2015} to show that core creation is able to reproduce the observed kinematics of low-mass dwarfs in the Local Group (see their Fig. 2).

Testing core creation as a solution of small-scale cosmological problems is not straightforward as the predictions of hydrodynamical simulations depend to a large extent on the implementation of subgrid prescriptions for baryonic physics \citep{Kim2014,SomervilleDave2015}. For this reason, it is often difficult to establish whether a proposed baryonic solution to a cosmological problem is a genuine one or rather just a consequence of the wide discretion in the implementation of subgrid physics. As far as core creation in particular is concerned, there are two promising ways to disentangle predictions that depend primarily on the implementation of baryonic physics from those that depend on the adopted cosmological model. The first is to obtain reliable kinematic measurements for field dwarf galaxies with extremely low stellar masses ($M_\ast \lesssim 10^6 \; M_\odot$). In such systems the DM density profile is mostly determined by the cosmological model since the available energy from stellar feedback is not sufficient to cause significant modifications \citep[e.g.,][]{diCintio2014a}. The second is to measure the kinematics of dwarfs at large galactocentric radii where the influence of the core is small. Since cores have typical sizes of $\sim 1$ kpc in dwarf galaxies \citep{PontzenGovernato2012}, the DM density profile at $R \gtrsim 2$ kpc is mostly determined by the cosmological model. 

In this article, we follow the second approach described above to test core creation in hydrodynamical simulations. We use the NIHAO simulation as an example, and repeat the comparison of \citet{Dutton2016} between the kinematics of low-mass halos and the kinematics of observed dwarfs. This time, however, the observational comparison sample consists of dwarfs with extended kinematics measured in the 21cm line of atomic hydrogen (HI). The article is organized as follows. In Section \ref{sec:datasets} we briefly describe the NIHAO hydrodynamical simulation and the observational sample of galaxies with HI kinematics used perform the kinematic comparison. In Section \ref{sec:method} we describe in detail the comparison setup. In Section \ref{sec:results} we present our result (Figure \ref{fig:nihao_vs_gals}), and discuss its scientific interpretation. In the same section we mention the assumptions under which our interpretation is valid, and elaborate on the main caveats associated with these assumptions. We conclude by summarizing our work in Sect. \ref{sec:summary}.

\section{Datasets}
\label{sec:datasets}

Halos are obtained from the NIHAO hydrodynamical simulation in the \LCDM \ cosmological context \citep{Wang2015}. NIHAO is a set of 100 zoom-in hydrodynamical simulations of {isolated} halos in the Planck one-year cosmology \citep{PlanckXVI}. The halos span a very wide range of masses (reaching all the way down to $M_{200} \sim 3 \times 10^9 \; M_\odot$) and are selected in an unbiased way in terms of halo concentration, spin parameter, and assembly history.
The resolution is chosen such that the mass profile of each halo can be well resolved down to $\sim 1\%$ of the virial radius, which is the relevant length scale for probing the observable properties of the hosted galaxies. Two common-resolution simulations of each halo are performed in NIHAO: The first is a dark matter only (DM-only) simulation using the \texttt{pkdgrav} code \citep{Stadel2001}. The second is a smooth particle hydrodynamics (SPH) simulation using the \texttt{gasoline} code \citep{Wadsley2004,Keller2014}. In the hydrodynamic run, subgrid baryonic physics have been implemented according to the MaGICC prescription \citep{Stinson2013}.  

Our observational sample of {field} 
galaxies consists of objects with HI interferometric observations compiled from the literature by \citet{PapastergisShankar2016}, with the addition of nine galaxies from \citet{Lelli2014}. The final sample consists of a total of 209 galaxies, spanning a very wide range of galaxy masses. Out of this parent sample, we use 31 low-mass dwarfs for the comparison with the NIHAO simulation (see Sec. \ref{sec:method}). Since these dwarfs are located at typical distances of $\lesssim 10$ Mpc, we also refer to our dwarf sample as the ``Local Volume'' sample of dwarfs. We note  that not all dwarfs in our sample are well-resolved enough to have a reliable measurement of their full rotation curve (RC). In fact, 23 out of the 31 dwarfs entering the comparison only possess a single kinematic measurement at the outermost HI point, \Vout $= V(R_\mathrm{out,HI})$.

In this article, we compare simulations of isolated halos with a sample of gas-rich galaxies that are located mainly in field environments. This choice allows us to evaluate more directly the potential cosmological relevance of the TBTF problem, as our comparison is not influenced by complex environmental effects such as tidal stripping and ram-pressure gas removal \citep[e.g.,][]{Zolotov2012,Arraki2014}. In addition, studying the TBTF problem for field galaxies leads to a more cosmologically representative assessment  because, in a volume-complete sense, more than 70\% of halos and galaxies are non-satellite objects \citep{Klypin2011,Karachentsev2013}.

\section{Method}
\label{sec:method}

\subsection{Matching observed dwarfs  with their NIHAO halos}

We match galaxies with NIHAO halos based on their rotational velocities derived from the spectral width of their HI line profiles, \Vrot. For galaxies, the profile of their HI emission line can be obtained from spectral radio observations, and the observed width of the profile at 50\% peak intensity level can be readily measured, $W_\mathrm{50,obs}$ (see Fig. 1 in \citealp{Papastergis2015}). The linewidth-derived velocity is then calculated from the inclination-corrected width of the profile as \Vrot $= W_\mathrm{50,obs}/(2 \times \sin i)$, where $i$ is the inclination of the galactic HI disk. For the NIHAO halos, \citet{Maccio2016} have derived simulated HI spectral profiles based on gas particle information in their hydrodynamic runs (see their Fig. 1). Since the simulated profiles have been extracted for an edge-on orientation of the halos, the linewidth-derived rotational velocity  in this case is simply \Vrot$ = W_\mathrm{50,edge}/2$.

We select dwarfs from our sample with linewidth-derived rotational velocities within the interval \Vrot$= 10 - 25$ \kms. According to \citet{Papastergis2015}, this is the relevant rotational velocity range for the field TBTF problem. We then compare the kinematics of  these dwarfs to NIHAO halos hosting simulated galaxies within the same \Vrot$= 10 - 25$ \kms \ interval. Keep in mind that \Vrot \ is used to match real dwarfs with their simulated counterparts in NIHAO; however, it is not used to carry out the kinematic comparison between the two  because \Vrot \ is a quantity computed from spatially {unresolved} spectral data of the HI line emission. The value of \Vrot \ does not therefore correspond to the rotational velocity at any specific galactocentric radius, and does not directly reflect any property of the host halo. In this respect, matching galaxies to their host halos in a simulation by means of \Vrot \ is conceptually similar to matching based on baryonic mass\footnotemark{}.

\footnotetext{In terms of baryonic mass, the \Vrot $= 10 -25$ \kms \ interval corresponds to the approximate range $M_\mathrm{bar} \approx 10^{6.5} - 10^8 \; M_\odot$. {Stellar} mass is not an appropriate quantity for matching purposes here since most of our field dwarfs are gas-dominated.
}

\subsection{Comparing the internal kinematics of dwarf galaxies and their host NIHAO halos}

Once we have matched real and simulated dwarf galaxies based on their linewidth-derived rotational velocities, \Vrot, we can   compare the internal kinematics of the two classes of objects. For the observed dwarfs, we consider the measured rotational velocity at their outermost HI radius, \Vout $= V(R_\mathrm{out,HI})$. It should be kept in mind that, unlike \Vrot, a measurement of \Vout \ requires {spatially resolved} observations of the HI line emission. This requirement means that our comparison must use the few tens of dwarfs with available HI interferometric data, rather than the hundreds of dwarfs detected in large area HI surveys performed with single-dish radiotelescopes \citep{Zwaan2004,Haynes2011}. In the case of NIHAO, we consider instead the circular velocity profiles published in \citet[their Fig. 3]{Dutton2016}. The circular velocity profiles of NIHAO dwarfs are extracted from the hydrodynamic runs of the simulation, and thus incorporate all baryonic effects on the matter content and mass profile of small halos (e.g., baryon depletion, core creation).

Since we compare observed HI kinematics with circular velocity profiles calculated from the enclosed mass in the simulation, we try to ensure that the \Vout \ values of the observed dwarfs correspond as closely as possible to the local value of circular velocity, \Vout \ $ = V_\mathrm{c}(R_\mathrm{out,HI})$. We therefore always apply corrections for pressure support to the measured \Vout \ values for our sample of observed dwarfs (see item \ref{item:complex_HI} in \S\ref{sec:caveats} for details on pressure-support corrections, and for a discussion of the important remaining caveats associated with our comparison between observed and simulated kinematics).

\section{Results and discussion}
\label{sec:results}

\begin{figure*}
\centering
\includegraphics[width=\linewidth]{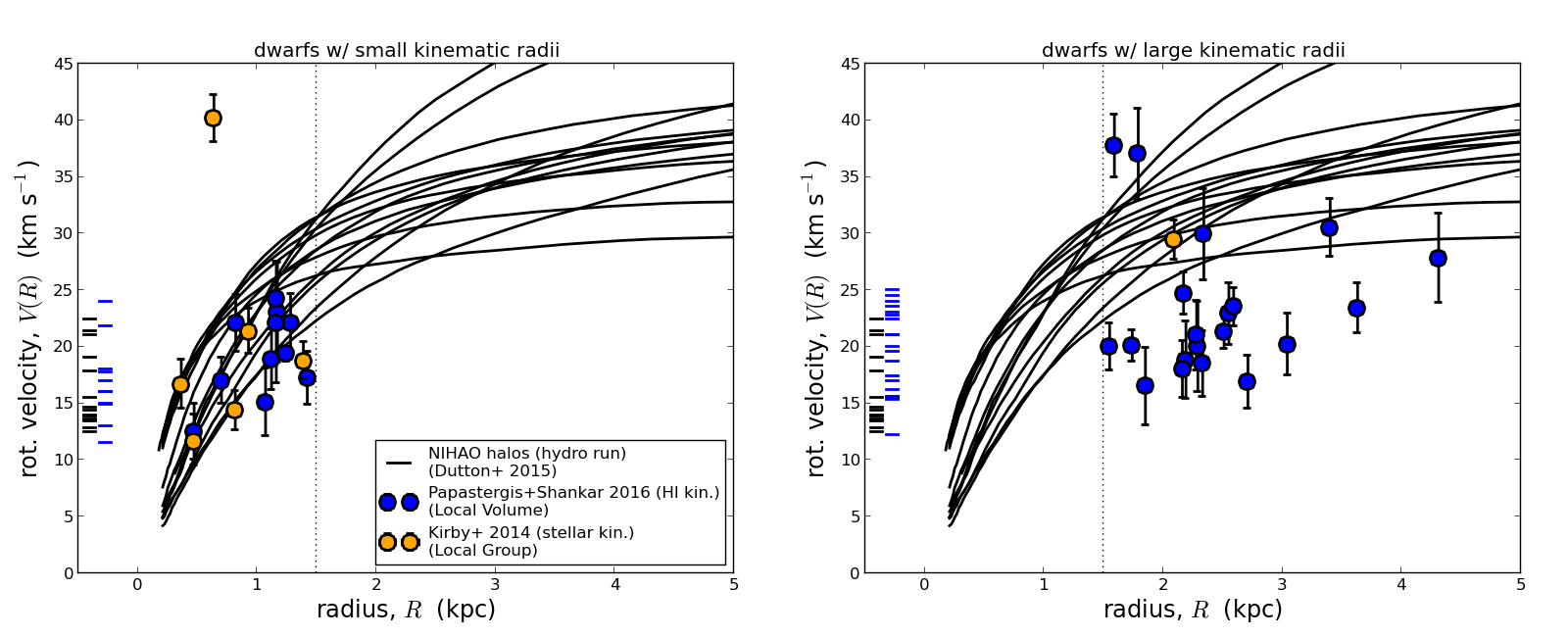}
\caption{
Comparison between the rotation curves of halos in the NIHAO hydrodynamical simulation and the HI kinematics of field dwarfs. 
Observed and simulated dwarfs are matched by requiring that their HI linewidth-derived rotational velocities are within the \Vrot $= 10 - 25$ \kms \ interval (individual \Vrot \ values are denoted respectively by blue and black tickmarks near the left $y$-axis). 
The black solid lines are total mass (i.e., dark+baryonic) circular velocity profiles for NIHAO dwarfs, extracted from the hydrodynamic runs of the simulation. The blue data points are 
instead rotational velocities measured at the outermost HI radius for observed dwarfs, corrected for pressure support. The orange data points are stellar kinematic measurements at the half-light radius for a sample of isolated dwarfs within the Local Group \citep{Kirby2014}.
The only difference between the left and right panel is that the former shows objects with small kinematic radii (\rout$ < 1.5$ kpc), while the latter shows objects with larger kinematic radii (\rout$ > 1.5$ kpc). See  Sect. \ref{sec:results} for the scientific interpretation of this figure.  
}
\label{fig:nihao_vs_gals}
\end{figure*}

Figure \ref{fig:nihao_vs_gals} shows the comparison between the internal kinematics of NIHAO halos and observed dwarfs with resolved HI data. We note  that RCs for the simulated dwarfs are obtained from the hydrodynamic runs of NIHAO, and therefore include all baryonic feedback effects on the mass profiles of low-mass halos. As Figure \ref{fig:nihao_vs_gals} shows, NIHAO dwarfs with very low values of linewidth-derived rotational velocity, \Vrot$= 10 - 25$ \kms, are hosted by relatively massive halos. The hydrodynamic RCs reach  typical peak velocities\footnotemark{} that are much higher, \VhaloHydro$= 35 - 40$ \kms. \citet{Maccio2016} show that this assignment of simulated dwarfs to relatively massive hosts allows the NIHAO simulation to successfully reproduce the dearth of dwarfs with low \Vrot \ values observed in the nearby universe \citep{Klypin2015}. Although they are relatively massive, however, NIHAO halos are also able to reproduce the low rotational velocities measured at $< 1.5$ kpc for nearby dwarfs (left panel of Fig. \ref{fig:nihao_vs_gals}). This holds true for both Local Group dwarfs with stellar kinematic measurements \citep{Kirby2014} and  for HI dwarfs with small kinematic radii. \citet{Dutton2016} attributes  this excellent agreement between the NIHAO predictions and the observed rotational velocities of dwarfs at small radii to core creation. More specifically, the presence of low-density cores in the central regions of halos leads to suppressed velocity profiles compared to the DM-only expectation (see their Fig. 3).

\footnotetext{For low-mass halos, the peak rotational velocity reached by a halo's RC in the hydrodynamic run is lower than that reached in the corresponding DM-only run, \VhaloHydro$<$\VhaloDM. This is due to the effect of baryon depletion \citep[e.g.,][]{Sawala2015}, whereby almost all the baryonic content of a low-mass halo is ejected by stellar feedback \citep[e.g.,][]{Wang2016}.}

However, the situation is strikingly different in the right panel of Figure \ref{fig:nihao_vs_gals}. The NIHAO hydrodynamic RCs fail to reproduce the observed HI rotational velocities of dwarfs with relatively large kinematic radii, \Rout \ $\sim 1.5 - 4$ kpc. Over this range of radii, the halo RCs reach velocities close to their peak value, while the HI rotational velocities measured for dwarfs remain low (\Vout \ $\lesssim 25$ \kms). The fundamental reason behind the result of Fig. \ref{fig:nihao_vs_gals} is that core creation can alter the mass profile of halos only within the innermost $1 - 1.5$ kpc. This is the physical region over which intense star-formation activity takes place in dwarf galaxies, and where the orbits of dark matter particles are subject to large and sudden potential fluctuations as a result \citep[see Fig. 4 in][]{PontzenGovernato2012}. At larger radii the kinematic effect of a core is small, and the velocity profile of a halo is primarily set by the underlying cosmological model. 

Our result illustrates the importance of testing the predictions of hydrodynamical simulations against a broad range of observational data, including field dwarfs with extended kinematic measurements. Such extended kinematic measurements are possible at present only through interferometric observations of the HI velocity fields of gas-rich objects. On the other hand, stellar kinematic measurements are confined to the optical half-light radius, which is typically $R_{1/2} \lesssim 1$ kpc in small dwarfs \citep{Kirby2014}. Consequently, samples of dwarfs with stellar kinematic measurements are not optimal for cosmological tests because they cannot probe the halo RC beyond the core region.

The results of this work and their interpretation are consistent with the findings of the closely related analysis performed in \citet{PapastergisShankar2016}. However, this previous work used a simplified approach in incorporating the effects of baryonic feedback on the kinematics of field dwarfs. For example, the effect of core creation was taken into account by employing the feedback-motivated ``DC14'' analytical mass profile \citep{diCintio2014b}. 
In order to fully specify this parametrized profile, a number of simplifying assumptions were required. For example, all halos were assigned median concentrations and differences in other physical properties among halos (e.g., spin) were not considered. The present analysis is more complete as the NIHAO halos span a wide range of properties such as concentration, spin, and assembly history.


Lastly, we would like to note that one possible concern regarding the result shown in Fig. \ref{fig:nihao_vs_gals} is that the HI linewidth, \Vrot, and the outermost HI point velocity, \Vout, are correlated to some extent (even though they are observationally distinct quantities). This creates the possibility that selecting dwarfs with small values of \Vrot \ may bias the plotted sample to objects that also have small values of \Vout, favoring objects with kinematics that do not agree with the simulation predictions. Even though we acknowledge the possibility that the partial correlation between \Vrot \ and \Vout \ may accentuate the disagreement between data and model in Fig. \ref{fig:nihao_vs_gals}, we believe that our results are robust. In particular, we have verified that matching dwarf galaxies and NIHAO halos in terms of baryonic mass --specifically, by demanding that their baryonic mass lies in the interval $M_\mathrm{bar} = 10^{6.5} - 10^8 \; M_\odot$ -- still reveals the presence of a kinematic discrepancy. Nonetheless, dwarfs selected based on baryonic mass display significant scatter in their kinematic properties in excess of the scatter predicted by NIHAO (see also  Appendix A in \citealp{PapastergisShankar2016}). As a result, the discrepancy in the case of baryonic mass selection is not as marked as shown in Fig. \ref{fig:nihao_vs_gals}.

\subsection{Assumptions and caveats}
\label{sec:caveats}

The result presented in Sec. \ref{sec:results} is only valid under certain assumptions, which are implicitly made in the analysis described above. In this section we would like to elaborate on the most important of these assumptions, and discuss under which conditions they may fail and therefore compromise the main scientific conclusions of this article.  

\begin{enumerate}[i]

\item \label{item:representative} \textit{Cosmologically representative dwarf sample}. The scientific interpretation of Figure \ref{fig:nihao_vs_gals} assumes that our literature sample of HI dwarfs is representative of the overall population of field dwarfs. In other words, we have assumed that the majority of field dwarfs have HI disks that extended out to $R > 1.5$ kpc (as Fig. \ref{fig:nihao_vs_gals} suggests). 
In reality, there is no guarantee that this assumption is correct. Our sample consists of a compilation of objects targeted for interferometric observations by various literature studies. This means that sample selection is not based on a single well-defined set of selection criteria\footnotemark{}. The biggest worry related to sample selection is a possible overrepresentation 
of galaxies with large values of \Rout. This cannot be ruled out because galaxies with larger HI disks tend to be brighter in the HI-line and therefore tend to be preferred targets for HI interferometric observations. If dwarfs with higher values of \Rout \ were outliers in terms of their kinematic properties (e.g., very low-concentration halos), then the result shown in  Fig. \ref{fig:nihao_vs_gals} could be attributed to sample selection bias.  



In the near future, the next generation of blind HI interferometric surveys (e.g., APERTIF surveys with the Westerbork interferometer; \citealp{Verheijen2008}; WALLABY survey with the ASKAP interferometer; \citealp{Duffy2012}) will enable us to assess this potential bias by delivering homogeneously selected samples of dwarfs with resolved HI kinematics. Moreover, these next-generation HI surveys are expected to detect a fair number of objects similar to LeoP \citep{Giovanelli2013}, i.e., gas-rich field dwarfs with extremely low stellar masses. These objects are not efficient at creating cores, due to their extremely low stellar-to-halo mass ratio \citep[e.g.,][]{diCintio2014a}, and therefore can be used as a complementary observational test of core creation (see Sec. \ref{sec:intro}).

\item \label{item:observational_systematics} \textit{No large observational systematics in the measurement of HI kinematics}. When plotting the dwarf data points in Fig. \ref{fig:nihao_vs_gals}, we have assumed that the observational data provided in the original literature references do not suffer from significant observational systematics \citep[see also discussion in \S3.3 of][]{Papastergis2015}. The largest concern here is the possibility of large biases in the reported inclination values for low-mass dwarfs \citep{Oman2016}. 

In general, determining the kinematic inclination for the HI disks of low-mass dwarfs is a highly uncertain process \citep[see, e.g., the case of LeoP in][]{Bernstein2014}. Even though we acknowledge the possibility that systematic inclination biases could be causing an apparent discrepancy between theoretical predictions and observations, we believe that this scenario is not very likely. More specifically, explaining the discrepancy would require a bias whereby most dwarfs in our sample are oriented more face-on than is estimated observationally. We believe that this type of bias is rather unlikely. First, the plotted data points are gathered from several different samples that use different methods for estimating galactic inclinations (titled ring fits, axial ratios of HI emission, axial ratios of stellar light). It is unlikely that  these different methods all suffer from the same type of inclination bias. Second, randomly oriented disks are more likely to be observed in a high-inclination  rather than a low-inclination orientation. This creates a small natural bias to underestimate (rather than overestimate) galactic inclinations. Third, most observational systematics, such as beam smoothing of the HI emission or disk thickness, tend to produce rounder HI disks and lead to underestimates\footnotemark{} of the inclination.

\footnotetext{One possible exception here is the SHIELD sample, which is selected from the ALFALFA blind HI survey \citep{Haynes2011}. In the case of SHIELD, dwarfs with low inclinations may be favored  because the sensitivity of a blind HI survey improves with decreasing observed linewidth, $W_\mathrm{50,obs} = W_\mathrm{50,edge} \times \sin i$ (see Fig. 12 in \citealp{Haynes2011}).}

\item \label{item:complex_HI} \textit{Observed HI kinematics accurately reflect the true circular velocity profile}. The RCs of the NIHAO halos plotted in Fig. \ref{fig:nihao_vs_gals} represent circular velocity profiles, computed from the enclosed dynamical mass as $V_c(R)^2 = G \, M_\mathrm{dyn}(<R) \: / \: R$. As a result, the comparison with observed HI kinematics implicitly assumes that HI rotational velocities (up to appropriate corrections) faithfully trace the true underlying potential. 

One worry here is that the magnitude of ordered rotation for low-mass dwarfs becomes comparable to the observed velocity dispersion of HI gas ($\sigma_\mathrm{HI} \sim 8-10$ \kms). This means that the pressure support provided by the dispersive motion of HI cannot be ignored in these systems. For this reason, the plotted \Vout \ values in Fig. \ref{fig:nihao_vs_gals} include a correction for pressure support. For some dwarfs in our sample this correction has been already performed in the original literature reference \citep[see, e.g.,  \S 3.1.2. in][]{Oh2015}. For the rest we apply the following approximate pressure support correction: $V_\mathrm{out,HI} \rightarrow \sqrt{V_\mathrm{out,HI}^2 + 2\sigma_\mathrm{HI}^2}$ with $\sigma_\mathrm{HI} = 8$ \kms. This correction follows Eq. 2 in \citet{Lelli2014}, assuming that the outermost velocity measurement is performed at two times the scale length of the HI disk. 

However, the biggest concern is whether correcting the observed rotation for pressure support as described above is enough to recover the true circular velocity. In particular, the pressure support corrections that are typically applied in observational studies assume a system in approximate equilibrium. This may not be the case for low-mass dwarfs, where energy injection through stellar feedback may create a complex velocity field with large radial motions \citep{Valenzuela2007,Teyssier2013,Marasco2016+}. In addition, feedback may heat up the HI disk in the vertical direction, creating a thick disk that rotates significantly slower than  is expected based on the local $V_c$ value \citep{Maccio2016,Verbeke2016+}. As Fig. \ref{fig:nihao_vs_gals} shows, even a $\sim 10$ \kms \ systematic underestimation of $V_c$ for small dwarfs would be enough to explain the discrepancy between the observations and the NIHAO simulation.

In principle, a systematic underestimation of the true enclosed mass due to the complexity of HI motions is a simple and plausible solution to the TBTF problem for field galaxies. However, it  should be kept in mind that testing this solution observationally may be extremely challenging. The reason is that the true circular velocity profile of a halo is not measurable through observations. As a result, it is not clear at present how to observationally distinguish between systems whose HI kinematics do or do not trace faithfully the underlying potential. It should also be kept in mind that if HI kinematics are indeed biased tracers of the true circular velocity profile, most of the observational support for the presence of central cores in low-mass dwarfs can be cast into doubt \citep[e.g.,][]{Oh2011b,BrookdiCintio2015b,Oh2015,Katz2016}.

\end{enumerate}

\section{Summary}
\label{sec:summary}

In this article we compare the kinematics of small halos in the NIHAO hydrodynamical simulation \citep{Wang2015} with the observed kinematics of a literature sample of dwarfs with available HI interferometric data \citep{PapastergisShankar2016}. Both halos and observed galaxies are selected to have HI linewidth-derived rotational velocities in the range \Vrot$= 10 - 25$ \kms, which is the relevant range for the too big to fail (TBTF) cosmological problem \citep{Boylan2011,Ferrero2012,Tollerud2014,Garrison2014,Papastergis2015}.

We show that the RCs of NIHAO halos obtained in the hydrodynamic run can successfully reproduce the kinematics of dwarfs with small kinematic radii, \rout$\lesssim 1.5$ kpc (left panel of Fig. \ref{fig:nihao_vs_gals}). According to \citet{Dutton2016}, this success can be attributed to the creation of feedback-induced cores in the density profiles of NIHAO halos. At the same time, however, we show that the RCs of NIHAO halos cannot reproduce the observed rotational velocities for dwarfs with relatively large kinematic radii, \rout$\approx 1.5 -4$ kpc (right panel of Fig. \ref{fig:nihao_vs_gals}). At these larger galactocentric radii, the kinematic effect of core creation is minimal; as a result, the velocity profiles of NIHAO halos cannot deviate much from the form dictated by the underlying cosmological model. Our result highlights the importance of testing hydrodynamical simulations against a variety of observational samples, including field dwarfs with HI kinematic measurements extending beyond the core region.

We would like to stress that the results we present in Fig. \ref{fig:nihao_vs_gals} are only valid under a set of implicit assumptions. The first is that our sample of dwarfs with spatially resolved HI kinematics is representative of the overall population of field dwarfs. This may not be the case since our sample has been compiled from the literature and lacks a common set of selection criteria (item \ref{item:representative} in \S\ref{sec:caveats}). Second, we assume that no large observational systematics affect the kinematic measurements of our sample of HI dwarfs. The main worry in this context is the possibility of a systematic overestimation of galactic inclinations (item \ref{item:observational_systematics} in \S\ref{sec:caveats}). Third, the comparison between halo RCs and measured HI velocities assumes that the HI kinematics reflect (up to corrections) the true underlying halo potential. This may not be the case if, for example, the HI velocity field is very complex or if the HI disk is very thick (item \ref{item:complex_HI} in \S\ref{sec:caveats}). Assumptions \ref{item:representative} and \ref{item:observational_systematics} can be tested in the near future with the data provided by next-generation HI interferometric surveys. On the other hand, assumption \ref{item:complex_HI} may be very difficult to test in practice since the true underlying potential in which a dwarf galaxy is embedded is not an observable quantity.

\begin{acknowledgements}
We thank an anonymous referee for insightful comments. We would also like to thank Aaron Dutton and Andrea Maccio for promptly sharing  the NIHAO simulation data, as well as Michael Jones and Thijs van der Hulst for providing helpful feedback on this manuscript. \\
E.P. is supported by a NOVA postdoctoral fellowship at the Kapteyn Astronomical Institute. A.A.P. acknowledges financial support of the DAGAL network from the People Programme (Marie Curie Actions) of the European Union’s Seventh Framework Programme FP7/2007-2013/ (under REA grant agreement number PITNGA-2011-289313). A.A.P. also acknowledges the Leids Kerkhoven–Bosscha Fonds (LKBF) for travel support.
\end{acknowledgements}



\begin{thebibliography}{}



\bibitem[Arraki et al.(2014)]{Arraki2014} Arraki, K.~S., Klypin, A., More, S., \& Trujillo-Gomez, S.\ 2014, \mnras, 438, 1466 






\bibitem[Begum et al.(2008a)]{Begum2008a} Begum, A., Chengalur, J.~N., Karachentsev, I.~D., Sharina, M.~E., \& Kaisin, S.~S.\ 2008, \mnras, 386, 1667



\bibitem[Bernstein-Cooper et al.(2014)]{Bernstein2014} Bernstein-Cooper, E.~Z., Cannon, J.~M., Elson, E.~C., et al.\ 2014, \aj, 148, 35

\bibitem[Boylan-Kolchin et al.(2011)]{Boylan2011} Boylan-Kolchin, M., Bullock, J.~S., \& Kaplinghat, M.\ 2011, \mnras, 415, L40 


\bibitem[Brook \& Di Cintio(2015a)]{BrookdiCintio2015a} Brook, C.~B., \& Di Cintio, A.\ 2015, \mnras, 450, 3920 

 \bibitem[Brook \& Di Cintio(2015b)]{BrookdiCintio2015b} Brook, C.~B., \& Di Cintio, A.\ 2015, \mnras, 453, 2133 

\bibitem[Brooks \& Zolotov(2014)]{BrooksZolotov2014} Brooks, A.~M., \& Zolotov, A.\ 2014, \apj, 786, 87 







\bibitem[Cannon et al.(2011)]{Cannon2011} Cannon, J.~M., Giovanelli, R., Haynes, M.~P., et al.\ 2011, \apjl, 739, L22

\bibitem[Chan et al.(2015)]{Chan2015} Chan, T.~K., Kere{\v s}, D., O{\~n}orbe, J., et al.\ 2015, \mnras, 454, 2981 









\bibitem[Di Cintio et al.(2014a)]{diCintio2014a} Di Cintio, A., Brook, C.~B., Macci{\`o}, A.~V., et al.\ 2014, \mnras, 437, 415

\bibitem[Di Cintio et al.(2014b)]{diCintio2014b} Di Cintio, A., Brook, C.~B., Dutton, A.~A., et al.\ 2014, \mnras, 441, 2986 

\bibitem[Duffy et al.(2012)]{Duffy2012} Duffy, A.~R., Moss, A., \& Staveley-Smith, L.\ 2012, \pasa, 29, 202 


\bibitem[Dutton et al.(2016)]{Dutton2016} Dutton, A.~A., Macci{\`o}, A.~V., Frings, J., et al.\ 2016, \mnras, 457, L74









\bibitem[Ferrero et al.(2012)]{Ferrero2012} Ferrero, I., Abadi, M.~G., Navarro, J.~F., Sales, L.~V., \& Gurovich, S.\ 2012, \mnras, 425, 2817





\bibitem[Garrison-Kimmel et al.(2014)]{Garrison2014} Garrison-Kimmel, S., Boylan-Kolchin, M., Bullock, J.~S., \& Kirby, E.~N.\ 2014, \mnras, 444, 222


\bibitem[Giovanelli et al.(2013)]{Giovanelli2013} Giovanelli, R., Haynes, M.~P., Adams, E.~A.~K., et al.\ 2013, \aj, 146, 15

\bibitem[Governato et al.(2010)]{Governato2010} Governato, F., Brook, C., Mayer, L., et al.\ 2010, \nat, 463, 203

\bibitem[Governato et al.(2012)]{Governato2012} Governato, F., Zolotov, A., Pontzen, A., et al.\ 2012, \mnras, 422, 1231






\bibitem[Haynes et al.(2011)]{Haynes2011} Haynes, M.~P., Giovanelli, R., Martin, A.~M., et al.\ 2011, \aj, 142, 170 





\bibitem[Karachentsev et al.(2013)]{Karachentsev2013} Karachentsev, I.~D., Makarov, D.~I., \& Kaisina, E.~I.\ 2013, \aj, 145, 101

\bibitem[Katz et al.(2016)]{Katz2016} Katz, H., Lelli, F., McGaugh, S.~S., et al.\ 2016, arXiv:1605.05971 

\bibitem[Keller et al.(2014)]{Keller2014} Keller, B.~W., Wadsley, J., Benincasa, S.~M., \& Couchman, H.~M.~P.\ 2014, \mnras, 442, 3013 

\bibitem[Kelvin et al.(2014)]{Kelvin2014} Kelvin, L.~S., Driver, S.~P., Robotham, A.~S.~G., et al.\ 2014, \mnras, 444, 1647


\bibitem[Kim et al.(2014)]{Kim2014} Kim, J.-h., Abel, T., Agertz, O., et al.\ 2014, \apjs, 210, 14

\bibitem[Kirby et al.(2014)]{Kirby2014} Kirby, E.~N., Bullock, J.~S., Boylan-Kolchin, M., Kaplinghat, M., \& Cohen, J.~G.\ 2014, \mnras, 439, 1015

\bibitem[Klypin et al.(2011)]{Klypin2011} Klypin, A.~A., Trujillo-Gomez, S., \& Primack, J.\ 2011, \apj, 740, 102

\bibitem[Klypin et al.(2015)]{Klypin2015} Klypin, A., Karachentsev, I., Makarov, D., \& Nasonova, O.\ 2015, \mnras, 454, 1798





\bibitem[Lelli et al.(2014)]{Lelli2014} Lelli, F., Verheijen, M., \& Fraternali, F.\ 2014, \aap, 566, A71






\bibitem[Macci{\`o} et al.(2016)]{Maccio2016} Macci{\`o}, A.~V., Udrescu, S.~M., Dutton, A.~A., et al.\ 2016, arXiv:1607.01028

\bibitem[Madau et al.(2014)]{Madau2014} Madau, P., Shen, S., \& Governato, F.\ 2014, \apjl, 789, L17

\bibitem[Marasco et al.(in prep.)]{Marasco2016+} Marasco, A., et al.\ in prep.

\bibitem[Mashchenko et al.(2008)]{Mashchenko2008} Mashchenko, S., Wadsley, J., \& Couchman, H.~M.~P.\ 2008, Science, 319, 174






\bibitem[Navarro et al.(1997)]{NFW1997} Navarro, J.~F., Frenk, C.~S., \& White, S.~D.~M.\ 1997, \apj, 490, 493 








\bibitem[Oh et al.(2011)]{Oh2011b} Oh, S.-H., Brook, C., Governato, F., et al.\ 2011, \aj, 142, 24

\bibitem[Oh et al.(2015)]{Oh2015} Oh, S.-H., Hunter, D.~A., Brinks, E., et al.\ 2015, \aj, 149, 180



\bibitem[Oman et al.(2016)]{Oman2016} Oman, K.~A., Navarro, J.~F., Sales, L.~V., et al.\ 2016, arXiv:1601.01026 

\bibitem[O{\~n}orbe et al.(2015)]{Onorbe2015} O{\~n}orbe, J., Boylan-Kolchin, M., Bullock, J.~S., et al.\ 2015, arXiv:1502.02036



\bibitem[Papastergis et al.(2011)]{Papastergis2011} Papastergis, E., Martin, A.~M., Giovanelli, R., \& Haynes, M.~P.\ 2011, \apj, 739, 38


\bibitem[Papastergis et al.(2015)]{Papastergis2015} Papastergis, E., Giovanelli, R., Haynes, M.~P., \& Shankar, F.\ 2015, \aap, 574, A113

\bibitem[Papastergis \& Shankar(2016)]{PapastergisShankar2016} Papastergis, E., \& Shankar, F.\ 2016, \aap, 591, A58


\bibitem[Planck Collaboration et al.(2014)]{PlanckXVI} Planck Collaboration, Ade, P.~A.~R., Aghanim, N., et al.\ 2014, \aap, 571, A16

\bibitem[Pontzen \& Governato(2012)]{PontzenGovernato2012} Pontzen, A., \& Governato, F.\ 2012, \mnras, 421, 3464





\bibitem[Read et al.(2016)]{Read2016} Read, J.~I., Iorio, G., Agertz, O., \& Fraternali, F.\ 2016, arXiv:1601.05821

\bibitem[Rodriguez-Puebla et al.(2016)]{BolshoiP} Rodriguez-Puebla, A., Behroozi, P., Primack, J., et al.\ 2016, arXiv:1602.04813







\bibitem[Sawala et al.(2015)]{Sawala2015} Sawala, T., Frenk, C.~S., Fattahi, A., et al.\ 2015, \mnras, 448, 2941 

\bibitem[Sawala et al.(2016)]{Sawala2016} Sawala, T., Frenk, C.~S., Fattahi, A., et al.\ 2016, \mnras, 457, 1931

\bibitem[Somerville \& Dav{\'e}(2015)]{SomervilleDave2015} Somerville, R.~S., \& Dav{\'e}, R.\ 2015, \araa, 53, 51 

\bibitem[Stadel(2001)]{Stadel2001} Stadel, J.~G.\ 2001, Ph.D.~Thesis, 3657 

\bibitem[Stinson et al.(2013)]{Stinson2013} Stinson, G.~S., Bovy, J., Rix, H.-W., et al.\ 2013, \mnras, 436, 625





\bibitem[Teyssier et al.(2013)]{Teyssier2013} Teyssier, R., Pontzen, A., Dubois, Y., \& Read, J.~I.\ 2013, \mnras, 429, 3068

\bibitem[Tollerud et al.(2014)]{Tollerud2014} Tollerud, E.~J., Boylan-Kolchin, M., \& Bullock, J.~S.\ 2014, \mnras, 440, 3511

\bibitem[Tollet et al.(2016)]{Tollet2016} Tollet, E., Macci{\`o}, A.~V., Dutton, A.~A., et al.\ 2016, \mnras, 456, 3542








\bibitem[Valenzuela et al.(2007)]{Valenzuela2007} Valenzuela, O., Rhee, G., Klypin, A., et al.\ 2007, \apj, 657, 773

\bibitem[Verbeke et al.(in prep.)]{Verbeke2016+} Verbeke, R., et al.\ in prep.

\bibitem[Verheijen et al.(2008)]{Verheijen2008} Verheijen, M.~A.~W., Oosterloo, T.~A., van Cappellen, W.~A., et al.\ 2008, The Evolution of 
Galaxies Through the Neutral Hydrogen Window, 1035, 265 




\bibitem[Wadsley et al.(2004)]{Wadsley2004} Wadsley, J.~W., Stadel, J., \& Quinn, T.\ 2004, \na, 9, 137

\bibitem[Wang et al.(2012)]{Wang2012} Wang, J., Frenk, C.~S., Navarro, J.~F., Gao, L., \& Sawala, T.\ 2012, \mnras, 424, 2715

\bibitem[Wang et al.(2015)]{Wang2015} Wang, L., Dutton, A.~A., Stinson, G.~S., et al.\ 2015, \mnras, 454, 83

\bibitem[Wang et al.(2016)]{Wang2016} Wang, L., Dutton, A.~A., Stinson, G.~S., et al.\ 2016, arXiv:1601.00967






\bibitem[Zolotov et al.(2012)]{Zolotov2012} Zolotov, A., Brooks, A.~M., Willman, B., et al.\ 2012, \apj, 761, 71 


\bibitem[Zwaan et al.(2004)]{Zwaan2004} Zwaan, M.~A., Meyer, M.~J., Webster, R.~L., et al.\ 2004, \mnras, 350, 1210


\end{thebibliography}
\end{document}